\documentclass[conference]{IEEEtran}
\pdfoutput=1
\IEEEoverridecommandlockouts
\usepackage{cite}
\usepackage{amsmath,amssymb,amsfonts}
\usepackage{algorithmic}
\usepackage{graphicx}
\usepackage{textcomp}
\usepackage{xcolor}
\usepackage[font=small,skip=1pt]{caption}
\usepackage{verbatim}
\def\BibTeX{{\rm B\kern-.05em{\sc i\kern-.025em b}\kern-.08em
    T\kern-.1667em\lower.7ex\hbox{E}\kern-.125emX}}
\begin{document}

\title{Exploring Web Search Engines to Find Architectural Knowledge}

\author{\IEEEauthorblockN{
Mohamed Soliman\IEEEauthorrefmark{1},
Marion Wiese\IEEEauthorrefmark{2}, Yikun Li\IEEEauthorrefmark{1}, 
Matthias Riebisch\IEEEauthorrefmark{2},
and
Paris Avgeriou\IEEEauthorrefmark{1}}
\IEEEauthorblockA{\IEEEauthorrefmark{1}
Bernoulli Institute for Mathematics, Computer Science and Artificial Intelligence \\
University of Groningen, Groningen, The Netherlands \\
\{m.a.m.soliman, yikun.li, p.avgeriou\}@rug.nl}
\IEEEauthorblockA{\IEEEauthorrefmark{2}Department of Informatics,
Universität Hamburg, Germany\\
Email: \{wiese, riebisch\}@informatik.uni-hamburg.de}
\thanks{This work was supported by ITEA3 and RVO under grant agreement No. 17038 VISDOM (https://visdom-project.github.io/website).}
}

\maketitle

\begin{abstract}
Software engineers need relevant and up-to-date architectural knowledge (AK), in order to make well-founded design decisions. However, finding such AK is quite challenging.  One pragmatic approach is to search for AK on the web using traditional search engines (e.g. Google); this is common practice among software engineers. Still, we know very little about what AK is retrieved, from where, and how useful it is. In this paper, we conduct an empirical study with 53 software engineers, who used Google to make design decisions using the Attribute-Driven-Design method. Based on how the subjects assessed the nature and relevance of the retrieved results, we determined how effective web search engines are to find relevant architectural information. Moreover, we identified the different sources of AK on the web and their associated AK concepts.
\end{abstract}


\section{Introduction}
\label{sec:introduction}
Architectural knowledge (AK)
is crucial for software engineers to make architectural design decisions~\cite{BassBook2012}. For instance, knowledge about 
technologies or architectural patterns, including their benefits and drawbacks, is important to select an architectural solution for a design problem.
However, finding architectural knowledge (AK) is a challenging task~\cite{GortonICSA2017} for a number of reasons. First, AK resides in multiple heterogeneous \textit{AK sources}, such as technology documentation~\cite{GortonICSA2017}, issue tracking systems~\cite{Bhat2017AutomaticApproach}, and developer communities (e.g. Stack Overflow)~\cite{SolimanWicsa2015}. Thus there is no single source of AK that contains all required AK. 

Second, each source of AK contains different \textit{AK concepts} (e.g. design decisions, solution alternatives~\cite{Zimmermann2009}, or the benefits and drawbacks of architectural solutions~\cite{SolimanWicsa2015}).
For instance, developer communities contain predominantly general AK concepts~\cite{Tang2010}, such as the benefits and drawbacks of architectural solutions~\cite{SolimanICSA2017}. In contrast, issue tracking systems contain mainly design and reasoning AK concepts~\cite{Tang2010}, such as design decisions of existing systems~\cite{Bhat2017AutomaticApproach}. Thus, depending on the AK concept, one may need to look into a different AK source.

Third, all AK sources are characterized by a fast pace of change, and accelerating technology churn~\cite{HassanStackoverflow}. This makes it even harder for software engineers to find and analyze information within the different sources of AK.



One approach to facilitate finding AK is to
manually capture AK (i.e. search for AK, and codify it) from multiple sources, and subsequently structure and store it in a repository (e.g.~\cite{GortonWicsa2015}). This supports software engineers to directly find relevant AK concepts, without navigating through many sources of AK depending on the concept. However, manually capturing AK requires significant efforts to gather and keep knowledge up to date; this makes it 
an expensive means in industrial practice~\cite{Capilla:2016:YSA}.

A more pragmatic way to search for AK from different sources is to use web search engines (e.g. Google). Web search engines are commonly used by software engineers in their daily business to find technical solutions for problems~\cite{Xia2017WhatWeb}. Moreover, web search engines can provide access to multiple AK concepts, such as design decisions from an existing system (e.g. within open source systems~\cite{Bhat2017AutomaticApproach}), as well as descriptions of architectural solutions (e.g. within technology documentation~\cite{GortonICSA2017}). It is even possible to use web search engines to populate AK repositories~\cite{GortonICSA2017}. 

While web search engines can provide support to find AK, recent experiences show that web search engines return many irrelevant results when searching for AK~\cite{GortonICSA2017}. In fact, there is little to no empirical evidence about: a) which AK sources and AK concepts can actually be found by web search engines; and b) when and how web search engines can be helpful for practitioners.


 
Our main \textbf{goal} in this paper is to 
\textit{explore which AK sources and AK concepts are retrieved by web search engines, and to gauge the effectiveness of web search engines to find relevant AK during the architectural design process}.
To this end, we conducted an empirical study with 53 software engineers with different levels of experience. The subjects used the most popular web search engine (i.e. Google) to find relevant AK concepts when conducting the steps of the Attribute-Driven Design (ADD) method  \cite{KazmanDesigningSoftware2016}; ADD was chosen as it is one of the most popular architectural design processes in literature.
The study results in the following contributions:
\begin{itemize}
\item We empirically identified AK sources, that web search engines are able to find. Moreover, we associated each AK source with the AK concepts they contain, based on the perspective of software engineers. Furthermore, we determined possible correlations between AK concepts.
\item We created a corpus of empirically classified and evaluated web pages and their respective AK sources and AK concepts.
\item We measured and compared the effectiveness of web search engines to support software engineers during the execution of the ADD steps. Moreover, we determined the most relevant AK sources for each ADD step according to the evaluation of software engineers.
\item We identified the AK concepts that make web pages highly relevant for software engineers when making design decisions.
\end{itemize}

The rest of the paper is structured as follows: Section \ref{sec:background} provides a background on the ADD steps, while Section \ref{sec:studydesign} presents the research questions and study design. Sections \ref{sec:RQ1Results}, \ref{sec:CorrelationConcepts}, \ref{sec:RQ2} and \ref{sec:RelevanceConcepts} present the results per research question, Section \ref{sec:discussion} discusses our results and their implications to practitioners and researchers, and Section \ref{sec:threats} discusses threats to validity. Finally, Section \ref{sec:relatedwork} discusses some related work, and Section \ref{sec:conclusion} concludes the paper.




\section{Attribute Driven Design Steps}
\label{sec:background}
Kazman et al. \cite{KazmanDesigningSoftware2016} proposed a number of iterative steps within ADD to make architectural decisions. For the purposes of our study, we select three of these steps, which are the most \textit{information-intensive} \cite{ASI:ASI20197}: they require searching for architectural information to be conducted. We explain these three steps and the type of information that software engineers need to perform them.

 \textit{Identify design concepts}: In this step, alternative architectural solutions are identified for a design issue. For example, a software engineer might look for alternative broker technologies, which might fulfill system requirements and align with the system constraints. As an example, these alternative solutions for broker technologies could be RabbitMQ, Kafka and ActiveMQ. To perform this step, software engineers need to search for information regarding alternative solution options.
 
\textit{Select design concepts}: In this step, one architectural solution is selected from a list of alternative solutions (from the previous step). This is done by comparing alternative solutions with each other regarding their ability to fulfill functional requirements and quality attributes. For example, RabbitMQ, Kafka and ActiveMQ are compared regarding performance and reliability to decide on the most suitable broker technology. To perform this step, software engineers need to search for benefits and drawbacks of the alternative options, e.g. information about performance benchmarks and reliability features for each of the alternative solutions.

\textit{Instantiate architecture elements}: In this step, the selected architectural solution (from the previous step) is customized to match the system requirements. For instance, to achieve high availability, replication tactics need to be implemented. This is done by configuring the selected broker technology (e.g. Kafka) to add a specific number of replicated instances. To perform this step, software engineers need to search for information and experiences regarding technology features, and their abilities to implement architectural tactics \cite{KazmanDesigningSoftware2016}.

\section{Study design}
\label{sec:studydesign}
\subsection{Research questions}
To achieve our goal (see Section \ref{sec:introduction}), we ask the following research questions (RQs):

\noindent\textit{(\textbf{RQ1}) Which AK can web search engines support to find?}
\begin{itemize}
    \item \textit{(\textbf{RQ1.1})} \textit{Which AK sources can web search engines find?} 
    \item \textit{(\textbf{RQ1.2}) Which AK concepts are prominent within the found AK sources?}
\end{itemize}


As discussed in Section \ref{sec:introduction}, researchers have explored multiple different AK sources and their AK concepts. However, there is no comprehensive list of AK sources and AK concepts on the web. Thus, we ask RQ1.1 to determine possible sources of AK that exist on the web. Answering RQ1.1 can confirm the AK sources that are already known but it can also reveal new AK sources, which have not been previously explored. Moreover, we ask RQ1.2 to determine the AK concepts, which commonly appear in each of the found AK sources. 

\noindent\textit{(\textbf{RQ2}) Which AK concepts co-occur on the web?}

We ask RQ2 to determine possible relationships between AK concepts on the web. Answering RQ2 can help us determine if AK concepts appear together on the web similarly to their conceptual relationships in AK ontologies (e.g. architectural solutions have benefits and drawbacks \cite{SolimanWicsa2015}). This can support assessing whether the relations between concepts in existing AK ontologies are reflected on the AK found in the web.

\noindent\textit{(\textbf{RQ3}) How well do web search engines support software engineers in following the ADD steps?}
\begin{itemize}
    \item \textit{(\textbf{RQ3.1})} \textit{How effective are web search engines to find AK needed for performing the ADD steps?} 
    \item \textit{(\textbf{RQ3.2}) 
    Which AK sources have the biggest contribution on the effectiveness of web search engines for each of the ADD steps?}

\end{itemize}

Within each step of the ADD (see Section \ref{sec:background}), software engineers need to search for different types of architectural information. We ask RQ3.1 to quantitatively measure the effectiveness of web search engines to find relevant architectural information, and to determine how much web search engines can support software engineers during the different architectural design steps. Moreover, we ask RQ3.2 to determine the most useful sources of AK for each ADD step. 
Answering RQ3.2 can support prioritizing and directing our future research efforts to explore and capture AK, by focusing on certain AK sources, that yield the highest benefit to software engineers.

\noindent\textit{(\textbf{RQ4}) Which AK concepts make web pages more relevant for design decisions?}

When using web search engines to perform architectural design tasks, some AK concepts may increase the relevance of the corresponding web pages. We ask RQ4 to determine which AK concepts indeed can make a web page more relevant for specific architectural tasks than others.
Answering RQ4 could provide guidance on which AK is worth sharing on the web. 

\subsection{Overview on the research process}
To answer the RQs, we conducted an exploratory case study \cite{Runeson2012CaseExamples} with 53 software engineers (see Section \ref{sec:participants}) who used the most popular web search engine (i.e. Google) to perform the ADD steps (see Section \ref{sec:background}). The conducted case study is an embedded case study, where the ADD steps constitute our case and the executed search queries from the 53 software engineers are the units of analysis.

To collect data, we asked the 53 software engineers to solve six architectural design searching tasks (see Section \ref{sec:tasks}) using Google, where each searching task performs one of the three ADD steps \cite{KazmanDesigningSoftware2016}, as explained in Section \ref{sec:background}. For each searching task, the participants executed multiple queries in Google, and assessed the resulted web pages regarding two aspects: 1) The relevance of each web page to the searching task, and 2) The AK concepts which exist in each web page. Further details are presented in Section \ref{sec:procedure}.

To analyze the collected data, and answer the RQs, we used the following analysis methods: 
\begin{itemize}
\item \textit{Web pages classification}: To answer RQ1.1, RQ1.2 and RQ3.2, we classified collected web pages (retrieved by Google for each query) into their respective AK source using a semi-automated approach (see Section \ref{sec:classification}).
\item \textit{Descriptive statistics and correlations}: To answer RQ1.2, RQ2 and RQ 3, we used descriptive statistics and evaluated correlations between AK sources, AK concepts and relevance of web pages using Pearson $\tilde{\chi}^2$ test \cite{Pearson1900} (see Section \ref{sec:statistic}).
\item \textit{Effectiveness measurement}: To answer RQ3.1 and RQ3.2, we measured the effectiveness of Google using standard information retrieval metrics (see Section \ref{sec:measurement}).
\end{itemize}

\subsection{Participants of the case study}
\label{sec:participants}
The participants of the case study are 53 software engineers. 50 of the participants attended a software architecture master course at the University of Hamburg, and 3 additional software engineers volunteered to participate in the study. An overview on the industrial experience of the participants is presented in Table \ref{tab:practioners}. Additional information regarding the technical background of the participants is available online.

\begin{table}[]
\centering
\caption{Industrial experience of participants}
\label{tab:practioners}
\begin{tabular}{cc|cc}
\hline
\multicolumn{2}{c|}{\textbf{Software development}} & \multicolumn{2}{c}{\textbf{Software architecture}} \\
\textbf{\# Years}                                & \textbf{\# Participants}                             & \textbf{\# Years}                                & \textbf{\# Participants}                              \\ \hline
\textgreater10 Years                             & 4                                                    & \textgreater5 Years                              & 5                                                     \\
3-10 Years                                       & 13                                                    & 2-5 Years                                        & 9                                                     \\
\textless3 Years                                        & 36                                                    & \textless1 Year                                           & 39                                                     \\ \hline
\end{tabular}
\end{table}

\subsection{Searching tasks}
\label{sec:tasks}
To answer the RQs, each participant (see Section \ref{sec:participants}) solved three searching tasks, where each task corresponds to one of the three ADD steps (see Section \ref{sec:background}). 

To support the validity of our results, we have designed two searching tasks for each ADD step. Thus, we have in total six tasks, from which we have randomly assigned three tasks (one task per ADD step) to each participant. The six searching tasks are real design problems, which have been gathered based on interviews with practitioners in a previous study \cite{Soliman2018ImprovingCommunities} within the field of architectural knowledge. Table \ref{tab:tasks} presents a brief description of the six searching tasks, and their relationship to the ADD steps. A complete description of each task is available online.


\begin{table}[]
\centering
\caption{Architectural searching tasks}
\label{tab:tasks}
\begin{tabular}{p{17mm}lp{58mm}lp{100mm}}
\hline
\textbf{ADD step}                                  & \textbf{ID} & \multicolumn{1}{c}{\textbf{Task description}}                                                                                                                                                                        \\ \hline
Identify design concepts          & T1          & For a realtime dashboard, identify middleware technologies which scale to \textgreater 100k users                                                     \\ \cline{2-3} 
                                                   & T2          & A system needs to communicate with mobile apps. Identify JSON parsers for Java with high performance, considering license constraints.                                                                               \\ \hline
Select design concepts            & T3          & A system communicates with a knowledge base via publish/subscribe patterns. Compare interoperability and latency of RabbitMQ, Kafka, and ActiveMQ.                                                                   \\ \cline{2-3} 
                                                   & T4          & Compare three technology families for big data systems: data collector, message brokers, and ETL engines. Requirements are throughput of 15,000 events/sec and availability of 99.99\%.                              \\ \hline
Instantiate architecture elements & T5          & CRM apps communicate with other systems using Apache Camel and RabbitMQ. Search for technology features and components designs to determine mechanisms channeling, translation and routing, and deployment topology. \\ \cline{2-3} 
                                                   & T6          & An application exposes services to other apps. Search for best practices regarding service decomposition to achieve high cohesion and low coupling.                                                                  \\ \hline
\end{tabular}
\end{table}

\subsection{Case Study procedures}
\label{sec:procedure}

\subsubsection{Preparations before the study}
\label{sec:before}
To support the validity of our results, it is important to ensure that the participants have a clear understanding about the procedures of the study, as well as the searching tasks, and AK concepts. Therefore, the authors met with the participants in two sessions and explained the study procedures, as well as the assessment of web pages regarding their relevance and AK concepts. After the first session, each participant received a user-guide (available online) with details about the study and a video tutorial on how to perform the tasks during the study. At the beginning of the second session and before conducting the study, the authors made a demo on assessing the relevance and specifying AK concepts for an example web page. In this demo, the participants were asked to specify the relevance and AK concepts for this example web page and provide their input via a polling feature in a web-conferencing tool. The polling results were discussed with the participants to show them how to correctly assess the relevance and how to specify AK concepts for each web page e.g. by paying particular attention to the requirements and constraints of the tasks.

We provide below the definitions for the degrees of relevance, that participants could choose from (in a five-level Likert scale):
\begin{itemize}
    \item  \textit{Very High Relevance (VH)}: The web page discusses a similar problem to that of the task and contains useful information. The web page provides an answer to the searching goal, and helps with fulfilling more than one requirement of the task.
    \item  \textit{High Relevance (H)}: The web page addresses a similar problem to that of the task and contains useful information. The web page provides an answer to the searching goal, and helps with fulfilling one requirement of the task.
    \item  \textit{Medium Relevance (M)}: The web page addresses a different problem to that of to the task at hand, but it provides some relevant information to the task, which could be an answer to the searching goal. Nevertheless, the provided information does not match specifically the task's requirements.
    \item  \textit{Low Relevance (L)}: The web page contains information, which is only remotely relevant to solving the given task, but might help for refining the search.
    \item  \textit{No Relevance (N)}: The web page has nothing to do with the task. It has no relevant information.
\end{itemize}

Moreover, each participant specified certain AK concepts for each web page. The list of AK concepts has been derived from existing literature \cite{Zimmermann12,SolimanWicsa2015,Jansen2005} and is as follows:
\begin{itemize}
    \item  \textit{Solution description}: general information on an architectural solution. 
    \item  \textit{Solution alternatives}: multiple (alternative) architectural options for a certain design issue. 
    The architectural options could be listed in the text or as a comparison of different options.
    \item  \textit{Solutions benefits}: information about the advantages of certain architectural solutions. 
    \item  \textit{Solutions drawbacks}: information about the disadvantages of certain architectural solutions, even discouraging their application. 
    \item  \textit{Made design decisions}: explanation about the architecture of a specific system. This includes the description of an existing architectural design of a specific system, or the explanation about certain design decisions of a specific system.
    \item  \textit{Others}: other relevant architectural information. 
\end{itemize}

\subsubsection{Study execution}
\label{sec:execution}
We asked the participants to perform the tasks in the order given to them. To ensure that the sequence of tasks does not influence the study, we provided each participant with a different sequence of tasks (available online). Moreover, we asked each participant to perform at least three queries per task and to evaluate the top 10 Google results for each query, as most users do not assess more than the top 10 results on the first page.

To facilitate specifying the relevance and AK concepts for each web page, we developed a Google Chrome plugin (provided online), which offers a user interface and stores submitted relevance and AK concepts in a database.

The participants started the study in a synchronous web-conference meeting, where they were able to directly ask for clarification if they had any uncertainties. By the end of the web-conference, the participants continued solving the tasks independently, while using the provided plugin to specify the relevance and AK concepts for each web page.

\subsubsection{After the study}
\label{sec:after}
After solving all the tasks, the participants were asked to fill out an exit survey (available online). We asked the participants about their experience searching for architectural information using Google, as well as the complexity to analyze web pages regarding AK concepts.

\subsection{Web pages classification}
\label{sec:classification}
As a result of the study, we received 5175 web pages (with 2623 unique pages) from executing 477 queries in Google, where each web page is evaluated from a participant regarding its relevance and AK concepts. To answer RQ1.1 and support answering the other RQs, we analyzed the resulted unique web pages to determine the AK source (e.g. forum, blog) for each page. To achieve this, we followed a semi-automated approach using two main steps, which are explained in the following sub-sections.

\subsubsection{Automatically clustering URLs of web pages}
We executed a clustering algorithm \cite{Yin2014AClustering} on the list of URLs to determine initial clusters of web pages. We have decided on this algorithm due to its excellence in clustering short text, compared to other classical clustering and topic modeling algorithms (e.g. LDA). Before executing the clustering, we filtered the URLs regarding symbols and stop words, and differentiated between host name and path within a URL.

We started the clustering algorithm using a big number of clusters (i.e. 100 clusters), and then reduced the number of clusters gradually after checking the results to reach the best possible clusters of URLs. Starting with a  big number of clusters facilitated determining the commonalities between smaller groups of URLs, which could be later aggregated into a single cluster. After several iterations of clustering, we achieved the best results by having 31 clusters.

By executing the clustering algorithm on the 2623 unique URLs, it succeeded to split the URLs into 27 consistent clusters with 1280 URLs (e.g. one consistent cluster is all blog pages in Apache websites), and 4 inconsistent clusters (i.e. mixed of different AK sources) with 1343 URLs. This separation has been determined by inspecting samples of web pages of the URLs within each cluster.

\subsubsection{Manually classifying web pages}
To determine the AK source for each web page, we started by analyzing the 27 consistent clusters to determine dominant AK sources in each cluster. Manually classifying web pages in the consistent clusters was done by inspecting sample web pages from the cluster to determine the dominant AK source category of this cluster. As a result of this step, we identified 15 initial categories of AK sources, and classified the 1280 URLs within the consistent clusters.

Based on the 15 initial categories of AK sources, the first three authors manually categorized the rest of the URLs (1343 web pages from within the 4 inconsistent clusters) into their respective categories of AK sources. This has been done by inspecting each of the web pages. While manually inspecting the web pages, we ignored offline web pages, spam and web pages in languages other than English or German. 
Moreover, a cross-check validation has been conducted between the first and second authors, as well as between the first and third authors to ensure agreement on the classification. To ensure agreement between the authors, we merged the 15 initial categories into 9 categories.

As a result of this step, we identified categories of AK sources on the web (see Section \ref{sec:RQ1}). Moreover, we created a corpus of 2522 unique web pages, which are categorized based on their AK sources.

\subsection{Measurement of effectiveness}
\label{sec:measurement}
To answer RQ3.1 and RQ3.2, we measure the effectiveness of Google using two metrics: $Precision@k$ and \textit{Normalized Discount Cumulative Gain}($nDCG@k$), where $k$ is the maximum number of search results that are considered for evaluation. We considered $k$ from 1 to 10 in our evaluations.

$Precision@k$ ~\cite{Manning2008IntroductionRetrieval} is the ratio between the number of relevant web pages (low or medium or high or very high), by the number of retrieved web pages in the results.

The ranking and relevance of the retrieved web pages are important factors to assess the effectiveness of search engines. However, Precision does not consider the ranking and relevance of web pages. Therefore, we use $nDCG@k$ ~\cite{Manning2008IntroductionRetrieval}, which consider both the ranking and relevance of the retrieved web pages. $nDCG@k$ is a well known metric in information retrieval and has been used successfully in software engineering research (e.g. \cite{ForumSearch}).

The main idea of $nDCG@k$ ~\cite{Manning2008IntroductionRetrieval} is to compare the ideal ranking of web pages ($IDCG$) to the ranking retrieved from a search engine for a certain query ($DCG$). For example, consider a task for which two users execute a query: the first user rates the top three web pages with relevance 3, 3, 1 while the second user rates the top three web pages with relevance 2, 2, 1. Meanwhile, the ideal ranking for this task is 3, 3, 2 when evaluating the top three search results. The $nDCG@3$ will compare the rankings 3, 3, 1 and 2, 2, 1 against 3, 3, 2. 

To compare the rankings of queries with the ideal ranking, we divide the $DCG@k$ of each query with the $IDCG@k$ (the $DCG@k$ of the ideal ranking).
The ideal ranking is based on combining the individual rankings of web pages (based on their relevance to a task) from  all participants in the study.

In order to translate the importance of relevance and ranking into a metric, we calculate the $DCG@k$ for each ranking (from each query). The $DCG@k$ provides different weights for web pages based on their relevance and ranking in the list of results (i.e. the higher the relevance and rank, the more weight). One common way to implement the $DCG@k$ is to use the logarithmic scale to provide the right weight based on the relevance and ranking. Specifically, we calculate $DCG@k = \sum_{i=1}^{k} \dfrac{2^{rel_{i}}-1}{log_{2}(i+1)} $, where $rel_{i}$ is the degree of relevance of a web page found by a query (based on the Likert scale defined in Section \ref{sec:before}).

To calculate the $nDCG@3$ for the previous examples q1 and q2, we calculate the $DCG@k$ for the rankings 3, 3, 1 (ranking of q1), 2, 2, 1 (ranking of q2) and 3, 3, 2 (the ideal ranking). 

\subsection{Statistical analysis and significance tests}
\label{sec:statistic}
To answer RQ1.2, RQ2 and RQ4 we used descriptive statistics. Moreover, we executed Pearson $\tilde{\chi}^2$ correlation tests \cite{Pearson1900} to determine the most relevant relationships.

To compute Pearson $\tilde{\chi}^2$ correlation tests and measure the phi coefficient we used the statistics tool SPSS\footnote{https://www.ibm.com/de-de/analytics/spss-statistics-software}.
Due to the large sample size, all correlations are significant (even to the 1\% level), but only the phi coefficients measured for RQ2 revealed correlations with effect sizes greater than 0.3 which means medium strong or strong correlations \cite{CohenStatisticalScienceDirect}.

To answer RQ1.2, we used descriptive statistics to determine relationships between AK sources (based on our classification see Sections \ref{sec:classification} and \ref{sec:RQ1}), and the AK concepts (as specified by practitioners, see Section \ref{sec:before}). For this, we counted the occurrences of each AK concept on the web pages and aggregated this for each AK source separately. 

To answer RQ2, we performed a two-tailed $T$-test ~\cite{Dean2017DesignExperiments} with $\alpha=0.05$ on all values of $Precision@k$ and $nDCG@k$ for each of the three ADD steps separately. Executing the $T$-test requires determining if the calculated values of $Precision@k$ and $nDCG@k$ have equal or unequal variances. Thus, we executed an $F$-test~\cite{Dean2017DesignExperiments} with $\alpha=0.05$ to determine if the values of $Precision@k$ and $nDCG@k$ have equal or unequal variances. Based on the results of $F$-test, we have executed the right method of $T$-test, either with equal or unequal variances.

We answered RQ4 using descriptive statistics to present the distribution of relevance within each AK concept, i.e. we evaluated only the web pages that contain the respective AK concept. Both the AK concepts and relevance of web pages have been specified by the participants during the study (see Section \ref{sec:before}). 

\section {RQ1: Architecture knowledge sources and concepts on the web}
\label{sec:RQ1Results}
\subsection{RQ1.1: AK sources on the web}
\label{sec:RQ1}
Table \ref{tab:AKsources} shows the identified AK sources, which have been retrieved by Google when searching for architectural information. Moreover, Table \ref{tab:AKsources} shows the percentages of each category based on our corpus (2522 web pages). \textbf{The results suggest that the predominant categories are blogs and tutorials as well as technology vendor documentations} (and to a lesser extent scientific contents). During our manual classification (see Section \ref{sec:classification}), we found a wide variety of blogs, such as personal blogs (e.g. Martin Fowler's blog\footnote{martinfowler.com}) and technology blogs (e.g. SAP technology blog\footnote{blog.sap-press.com}). Moreover, technology vendor documentations come at different levels of detail, from a high level description of technology to detailed code specification (e.g Apache technologies\footnote{kafka.apache.org/documentation}\footnote{flink.apache.org}).

On the other hand, \textbf{source code repositories (e.g. Github) and knowledge repositories are not commonly found} in the search results, and they come in less variety. Furthermore, the results show that \textbf{issue tracking systems (e.g. Apache issue trackers\footnote{issues.apache.org}) are not retrieved at all by Google} when searching for architectural information.
\begin{table}[]
\centering
\caption{AK sources on the web}
\label{tab:AKsources}
\begin{tabular}{rll}
\hline
\multicolumn{1}{c}{\textbf{AK source}}                                      & \multicolumn{1}{c}{\textbf{Example}}                                                   & \multicolumn{1}{c}{\textbf{\%}} \\ \hline
Blogs and tutorials                                                         & dzone.com                                                                              & 39\%                            \\ \hline
\begin{tabular}[c]{@{}r@{}}Technology vendor \\ documentations\end{tabular} & metamug.com/docs                                                                       & 23\%                            \\ \hline
Scientific contents                                                         & ieeexplore.ieee.org                                                                    & 13\%                            \\ \hline
Forums                                                                      & stackoverflow.com                                                                      & 7\%                             \\ \hline
\begin{tabular}[c]{@{}r@{}}Technical books and \\ white papers\end{tabular} & livebook.manning.com                                                                   & 4.5\%                           \\ \hline
Source code repositories                                                    & github.com                                                                             & 4.5\%                           \\ \hline
Knowledge repositories                                                      & stackshare.io                                                                          & 4\%                             \\ \hline
Presentations and videos                                                    & slideshare.net                                                                         & 2.7\%                           \\ \hline
\begin{tabular}[c]{@{}r@{}}Others \\ (e.g. tools, patents)\end{tabular}     & \begin{tabular}[c]{@{}l@{}}google.com/patents\\ json.parser.online.fr\end{tabular} & 2.3\%                           \\ \hline
\end{tabular}
\end{table}
\setlength{\dbltextfloatsep}{5pt}
\setlength{\dblfloatsep}{5pt}

\subsection {RQ1.2: Prominent AK concepts in the AK sources}
\label{sec:ConceptsPerSource}

\begin{figure*}
\setlength{\belowcaptionskip}{-10pt}
	\centering
		\includegraphics[scale=0.55]{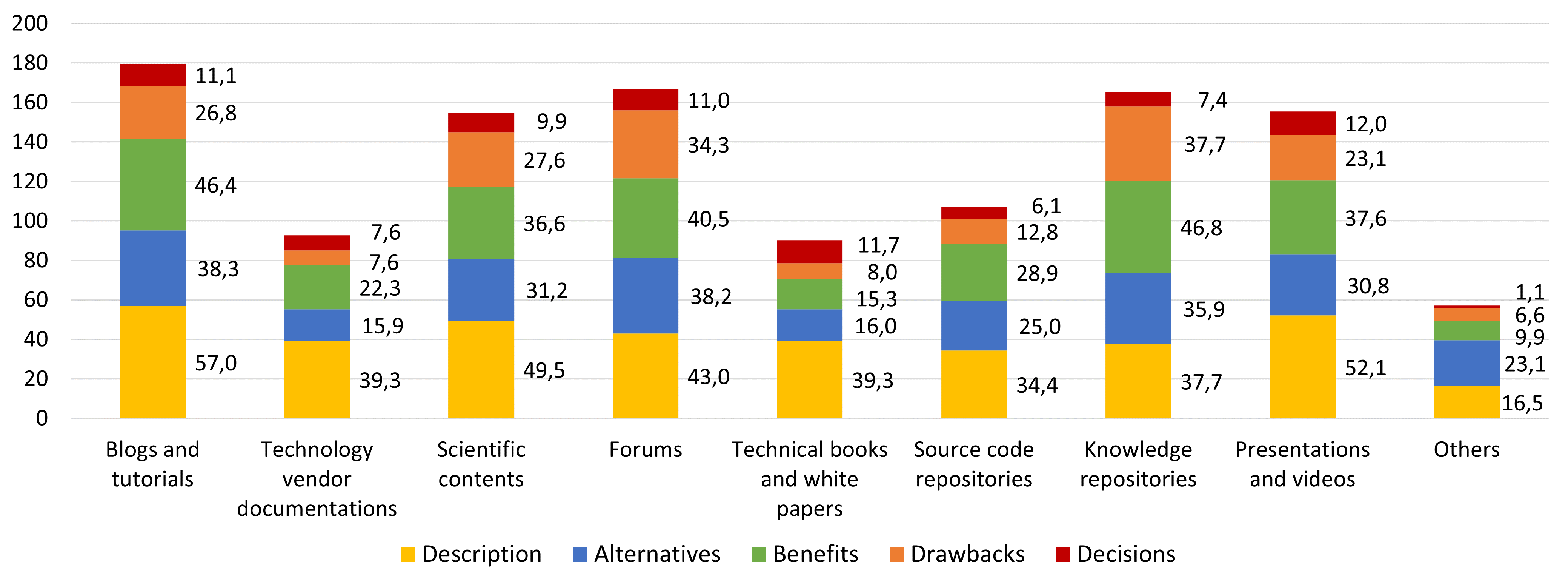}       
		\caption{Distribution of AK Concepts per AK source}
	\label{fig:ConceptsPerScource}
\end{figure*}

Figure \ref{fig:ConceptsPerScource} shows the percentages of occurrence for each of the AK concepts in the found AK sources (see Section \ref{sec:RQ1}). Because each web page can contain multiple AK concepts, the sum of percentages for all AK concepts in an AK source can exceed 100\% (for more details see the online resources.
Based on Fig. \ref{fig:ConceptsPerScource}, we can observe the following:
\begin{itemize}

\item Solution alternatives, benefits and drawbacks are less present within the technology vendor documentations and source code repositories; presumably this is because these AK sources usually discuss a single architectural solution, such as the documentation or source code of a specific technology.
\item \textbf{Made design decisions are underrepresented in all AK sources}. Thus, finding concrete examples of design decisions (e.g. a decision on specific components of a system), and their rationale for existing systems are not as easy to find among the AK sources on the web.
\item Solution descriptions are more prominent than the other AK concepts within technology vendor pages. This is probably because this AK source focuses on describing a certain architectural solution in more detail.

\end{itemize}

\section{RQ2: Co-occurrences of AK concepts}
\label{sec:CorrelationConcepts}
In order to evaluate the co-occurrences of AK concepts in web pages, we measured the correlation between the AK concepts (as explained in Section \ref{sec:statistic}).
Table \ref{tab:correlationConcepts} shows the correlation coefficients between each pair of AK concepts.

From Table \ref{tab:correlationConcepts}, we can observe that \textbf{the correlation between benefits and drawback stands out with a correlation coefficient of 0.651} which means a strong correlation exists\cite{CohenStatisticalScienceDirect}. This means that either a) web pages containing benefits often also contain drawbacks or b) web pages containing drawbacks often also contain benefits. To determine the right interpretation for the correlation between benefits and drawbacks, we inspected the exact frequencies of co-occurrences between benefits and drawbacks (see Table \ref{tab:BenefitsDrawbacks}). 
We can observe that drawbacks are rarely presented without benefits (46 out of 1152), while there are more web pages containing benefits without drawbacks (820 out of 1929). This means that 
\textbf{it is common to find web pages with benefits and no drawbacks, while it is rare to find web pages with drawbacks and no benefits}. This can be also seen in Fig. \ref{fig:ConceptsPerScource}. For example, 
blogs and tutorials contain nearly twice the amount of benefits compared to drawbacks. This indicates that software engineers in communities tend to praise the benefits of technologies rather than realistically evaluate the pros and cons of technologies equally.

From Table \ref{tab:correlationConcepts}, we can also observe that the correlations between alternatives and benefits, as well as between alternatives and drawbacks (0.355 and 0.340 respectively) are also medium strong\cite{CohenStatisticalScienceDirect}. This indicates that \textbf{lists of alternative solutions, are usually accompanied with a comparison between them regarding their benefits and drawbacks}.
One common example for this are knowledge repositories and forum entries containing alternatives, with their drawbacks and benefits.

A final observation from Table \ref{tab:correlationConcepts}, is a medium strong correlation (0.347) \cite{CohenStatisticalScienceDirect} between solution descriptions and benefits. This is quite common in technology vendor pages, where they commonly describe their technologies and their benefits, while omitting their drawbacks (see Fig. \ref{fig:ConceptsPerScource}).


\begin{table}[]
\centering
\caption{Correlation between each of the AK concepts}
\label{tab:correlationConcepts}
\begin{tabular}{l|c c c c c}
\hline
 & \textbf{Descr.} & \textbf{Altern.} &	\textbf{Benef.} & \textbf{Drawb.} & \textbf{Decisions}	\\
\hline
\textbf{Description}    & 1.000     & 0.181 & \textbf{0.347}    & 0.203             & 0.222 \\
\textbf{Alternatives}   &           & 1.000 & \textbf{0.355}    & \textbf{0.340}    & 0.079 \\
\textbf{Benefits}       &           &       & 1,000             & \textbf{0.651}    & 0.157 \\
\textbf{Drawbacks}      &           &       &                   & 1.000             & 0.122 \\
\textbf{Decisions}       &           &       &                   &                   & 1.000 \\
\hline
\end{tabular}
\end{table}

    \begin{table}
    \centering
    \caption{Co-occurrences between Benefits and Drawbacks}
    \label{tab:BenefitsDrawbacks}
        \begin{tabular}{c|r r|r}
        \hline
        	& \multicolumn{2}{c|}{\textbf{Drawbacks}} & \\
        \textbf{Benefits}	&  does not contain &  contains & \\
        \hline
        does not contain     & 3203     & 46      & 3249\\
        contains             &  820     & 1106    & 1929\\
        \hline
        \textbf{Sum}        & 4023      & 1152    & 5175\\
        \hline
        \end{tabular}
    \end{table}
    \setlength{\dbltextfloatsep}{5pt}
\setlength{\dblfloatsep}{5pt}
%


\section {RQ3: Search engines support for ADD steps}
\label{sec:RQ2}
\subsection {RQ3.1: Effectiveness of search engines to support ADD}
\label{sec:RQ2-A}

\begin{figure}
\setlength{\belowcaptionskip}{-10pt}
	\centering
		\includegraphics[scale=0.35]{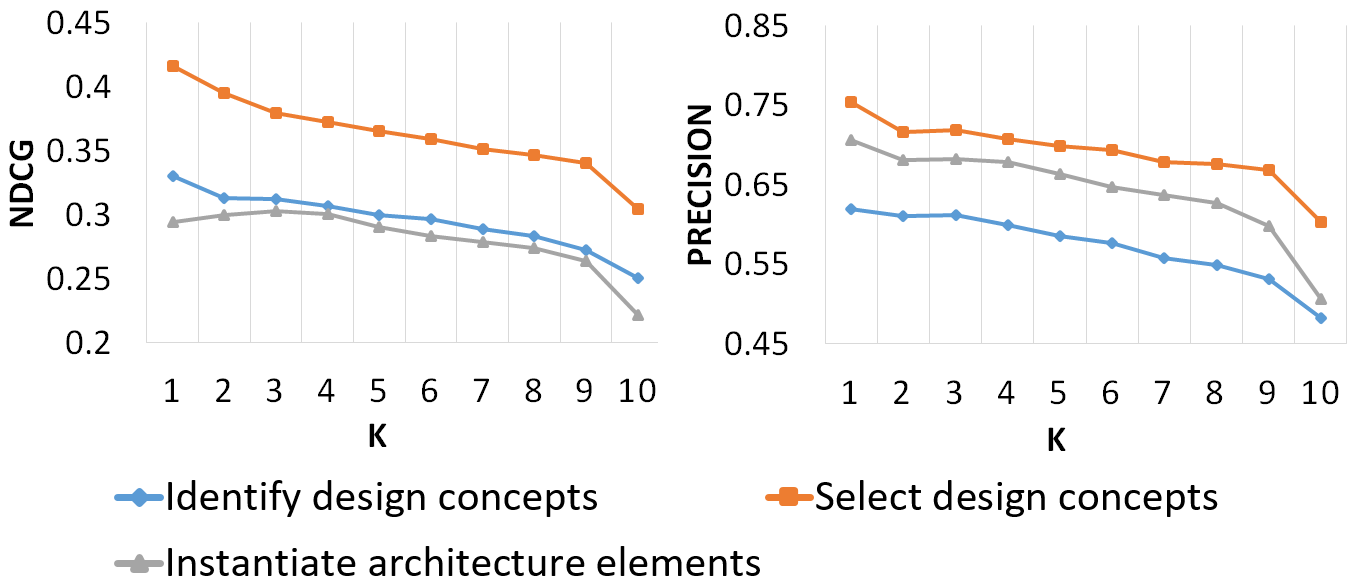}        
		\vspace*{1mm}
		\caption{Average nDCG and precision in finding architectural information for each ADD step.}
	\label{fig:averagePrecisionAndNDCG}
\end{figure}
\setlength{\dbltextfloatsep}{15pt}
\setlength{\dblfloatsep}{15pt}
Figure \ref{fig:averagePrecisionAndNDCG} shows the average $nDCG@k$ and $Precision@k$ when searching for architectural information during the three ADD steps.
From Fig. \ref{fig:averagePrecisionAndNDCG}, we can observe that the maximum average $Precision@1$ is 0.75 (for the ``Select design concepts'' step), which means that on average 75\% of the queries retrieved a relevant (low, medium, high, very high) web page at the top result (k=1) from Google. On the other hand, the lowest $Precision@10$ is 0.48 (for the ``Identify design concepts'' step), which means that on average 48\% of the retrieved top 10 web pages were relevant (i.e. have relevance of low, medium, high, very high); in this case, web search engines return both relevant and irrelevant results equally.
 
To compare the differences between ADD steps regarding their $nDCG@k$ and $precision@k$, we executed a significant $T$-test (as explained in Section \ref{sec:statistic}). The detailed results of the tests are available online.

From the test, \textbf{we found that the ``Select design concepts'' step has significantly higher $nDCG@k$ for k between 1 and 9 ($nDCG@k_{1\rightarrow 9}$) compared to the other two design steps}, while there is no significance difference in $nDCG@k_{1\rightarrow 10}$ between the ``Identify design concepts'' and ``Instantiate architecture elements'' steps. For example, from Fig. \ref{fig:averagePrecisionAndNDCG} the average $nDCG@1$ for the ``Select design concept'' ADD step is 0.42, while the average $nDCG@1$ is 0.29 for the ``Instantiate architecture elements'' ADD step. This means that on average 42\% of the ``Select design concept'' queries retrieved highly relevant (i.e. very high) web pages at the top result from Google, compared to just 29\% for the ``Instantiate architecture elements''. Thus, \textbf{finding AK to select an architectural solution from alternatives is easier than finding AK to instantiate a certain architectural solution}. 

The significance test also showed that the ``Select design concepts'' step has significantly better $precision1\rightarrow 9$ compared to the ``Identify design concepts''. In contrast, there is only a slight difference of $precision1\rightarrow10$ between the ``Select design concepts'' and ``Instantiate architecture elements''.


Looking at both the $nDCG@k$ and $precision@k$, on the one hand we can notice that the ``Identify design concepts'' has significantly lower $nDCG@k$ and $precision@k$ compared to the ``Select design concepts''. This means \textbf{it is harder to find AK on alternative architectural solutions  than to find information on how to compare them}.
On the other hand, the ``Instantiate architecture elements'' step has significantly lower $nDCG$ compared to the ``Select design concept'' step but a comparable $Precision$.
Because $nDCG$ considers the ranking and relevance of results compared to $Precision$, the $nDCG$ and $Precision$ of the ``Instantiate architecture elements'' indicate that \textbf{Google finds many distantly relevant web pages (e.g. general concepts on components design) to support the ``Instantiate architecture elements'' step}. In contrast, it is challenging for Google to find highly relevant solutions, which fulfill specific requirements.
\subsection{RQ3.2: The most influential AK sources}
\label{sec:RQ2.2}
\begin{figure*}
\setlength{\belowcaptionskip}{-10pt}
	\centering
		\includegraphics[scale=0.37]{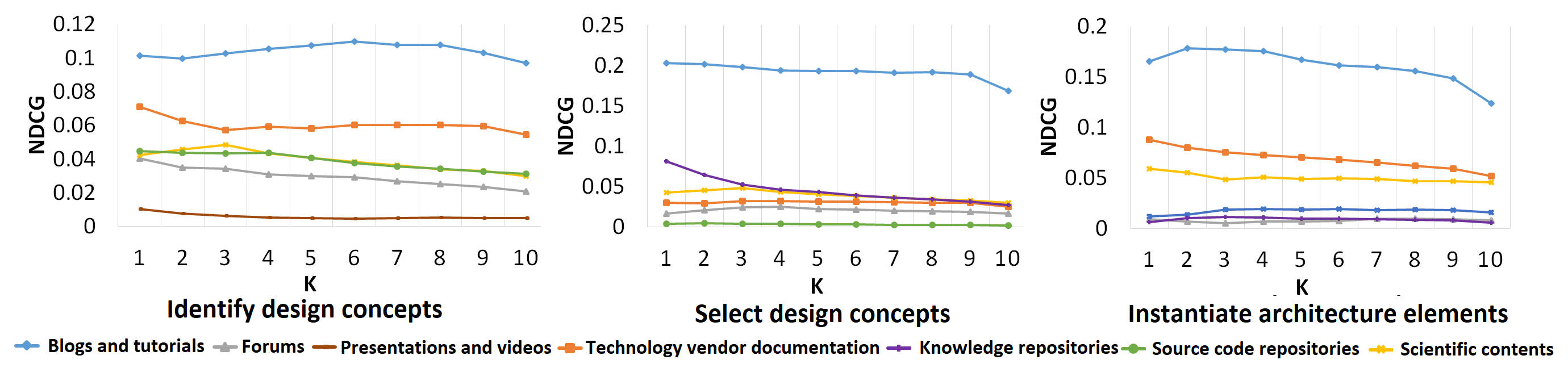}        
		\caption{The contribution of each AK source in the effectiveness of Google per ADD step}
	\label{fig:AKsourcesNDCG}
\end{figure*}
Figure \ref{fig:AKsourcesNDCG} shows the top six AK sources (see Table \ref{tab:AKsources}) with the highest effectiveness in each of the ADD steps. 
We can observe the following:

\textit{Blogs and tutorials} have the biggest contribution on the effectiveness of Google in all three ADD steps. Thus, \textbf{blogs have the highest relevance and highest ranking, and show to contain the most useful AK compared to other sources}.

\textbf{\textit{Technology vendor documentation} shows to have the second highest contribution on the effectiveness of Google for both the ``Identify design concepts'' and the ``Instantiate architecture elements'' steps}. This indicates their usefulness to identify options for architectural solutions. Moreover, technology vendor documentation contains detailed information regarding the technology features, which are useful for the ``Instantiate architectural elements'' step.
As shown in Section \ref{sec:ConceptsPerSource} solution alternatives and solution drawbacks are underrepresented in technology vendor documentation, which may be the reason why they are not as effective for the ``Select design concepts'' step.

\textbf{\textit{Knowledge repositories} (e.g. \cite{GortonWicsa2015}) show to be effective for $@k_{1\rightarrow 2}$ (i.e. for the top two Google results) only within the ``Select design concepts'' step}, while it has a negligible contribution on the effectiveness within the other two ADD steps. This is because most knowledge repositories on the web (e.g. stackshare) contain mainly high level information to compare multiple technologies with each other; they can thus somewhat help with ``Selecting design concepts''.

\textbf{\textit{Scientific contents} (e.g. papers or thesis) show to have a higher contribution on the effectiveness of Google than forums and source code repositories, for the ``Instantiate architecture elements'' step}. This could be due to the scarcity of AK regarding component designs in forums and source code repositories, while scientific contents are rich with component designs (e.g. reference architectures).

\textbf{\textit{Forums and source code repositories} show to have a limited influence on the effectiveness of Google of searching within the three ADD steps}. This might be because both forums and source code repositories are not well found by Google (see Table \ref{tab:AKsources}) compared to the other AK sources.

\section {RQ4: AK concepts in highly relevant web pages}
\label{sec:RelevanceConcepts}
\begin{figure}
\setlength{\belowcaptionskip}{-10pt}
	\centering
		\includegraphics[scale=0.5]{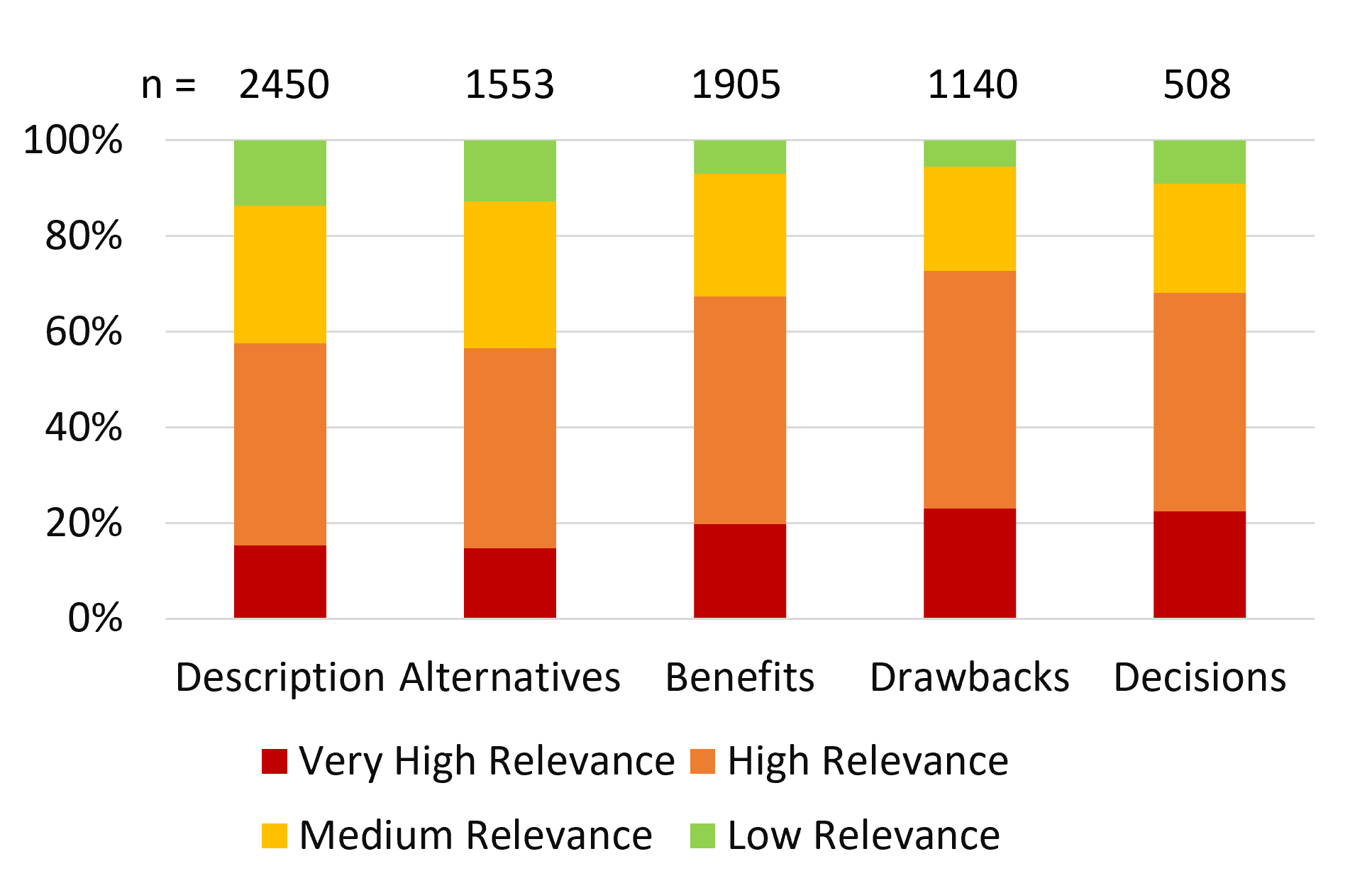}       
		\caption{Distribution relevance per concept}
	\label{fig:RelevanceOfAKconcepts}
\end{figure}
\setlength{\dbltextfloatsep}{15pt}
\setlength{\dblfloatsep}{15pt}
Figure \ref{fig:RelevanceOfAKconcepts} shows the total number of relevant web pages (presented as ``n'' in the figure), and the percentages of the different degrees of relevance (low, medium, high and very high as rated by the participants), in which AK concepts appear. 
From Figure \ref{fig:RelevanceOfAKconcepts}, we can observe that \textbf{web pages that contain solution benefits, solution drawbacks and made design decisions are most relevant for finding AK}, while web pages with solution description' and solution alternatives have a lower probability to be high or very high in relevance. For example, a web page containing solution drawbacks has a higher probability to be rated high or very high in relevance (72.7\%) than a web page containing solution alternatives (56.5\%). (Details are available online).

Even though, solutions drawbacks and made design decisions are not found by Google in its results as often as solution descriptions and solution alternatives, they showed to be the most common AK concepts in highly relevant pages, and thus they could be the most important for making design decisions.

\section {Discussion}
\label{sec:discussion}
\subsection{RQ1: AK sources and concepts on the web}

\subsubsection{Implications for practitioners}
The list of AK sources on the web (see Table \ref{tab:AKsources}) and their respective AK concepts (see Fig. \ref{fig:ConceptsPerScource}) could guide practitioners to determine the scope of searching (i.e. to search in the whole web or in specific web sites). For example, forums like Stackoverflow have shown to contain useful AK (e.g. \cite{Wicsa2016,SolimanICSA2017}) about the benefits and drawbacks between architectural solutions (see Fig. \ref{fig:ConceptsPerScource}). However, forums are not well considered by Google (see Table \ref{tab:AKsources}). Thus, practitioners should focus the scope of searching on specific forums like Stackoverflow to find more AK on the benefits and drawbacks between solutions. 


 

\subsubsection{Implications for researchers}
The list of AK sources in Table \ref{tab:AKsources} provides an overview of possible web sources of AK, from which researchers could extend current approaches to document AK (e.g. \cite{GortonICSA2017}). On the one hand, blogs are well indexed by Google but not previously explored by researchers for AK. Thus current studies on AK (e.g. \cite{Wicsa2016,SolimanICSA2017}) could be extended to explore the AK in blogs. On the other hand, some AK sources are previously explored by researchers for AK (e.g. forums \cite{Wicsa2016} and issue tracking systems \cite{Bhat2017AutomaticApproach}), but they are not well indexed by Google. These AK sources deserve extra attention to improve their ranking on web search results.

The distribution of AK concepts in each AK source in Fig. \ref{fig:ConceptsPerScource} show that there is no single AK source which contains specific AK concepts. However, certain AK concepts come more often in certain AK sources. Thus, researchers could make use of this information to develop AK documentation approaches, which consider multiple AK sources. For example, an AK approach can find and document benefits and drawbacks from forums, and solution description from technology vendor documentations.
\subsection{RQ2: Co-occurrences of AK concepts}
\subsubsection{Implications for practitioners}
The results in Section \ref{sec:CorrelationConcepts} show that benefits are preconditions for the occurrence of drawbacks in web pages, and that drawbacks rarely come alone in a separate web page. The results increase the awareness of practitioners that the predominant presence of benefits on the web does not mean lack of drawbacks for architectural solutions; it rather means the rarity of drawbacks on the web. To resolve this problem, practitioners should use web search engines to search explicitly for the drawbacks of architectural solutions.

\subsubsection{Implications for researchers}
The co-occurrences between AK concepts in Section \ref{sec:CorrelationConcepts} shows that AK concepts do not come all together in a single web page. However, subsets of AK concepts co-occur together more often than others. For instance the correlations between AK concepts in Table \ref{tab:correlationConcepts} could be grouped into three subsets of AK concepts: (benefits and drawbacks), (alternatives, benefits and drawbacks), and (Solution description and benefits). Researchers need to consider this division of AK concepts on the web, when developing approaches to automatically capture AK (i.e. search for AK, and codify it), e.g. developing dedicated AK capturing approaches for each subset of AK concepts.



\subsection{RQ3: How search engines support the ADD steps}

\subsubsection{Implications for practitioners}
The results in Section \ref{sec:RQ2-A} verified previous experiences with practitioners \cite{GortonICSA2017} that web-search engines return many irrelevant results. However, our study provides the first precise evidence on the effectiveness of Google to find AK for each ADD step. The differences in the effectiveness between the ADD steps (as shown in Fig. \ref{fig:averagePrecisionAndNDCG}) can help increase the awareness regarding the expected relevance of the retrieved results. This can guide practitioners to determine when to rely on web search engines, and when to seek other ways (e.g. asking experts) to search for AK. For instance, practitioners could rely on search engines to find AK for the ``Select design concepts'' step. But, they are better off seeking other ways during the ``Instantiate architecture elements'' step, as many of the retrieved results from web search engines are distantly relevant.


\subsubsection{Implications for researchers}
The results in Section \ref{sec:RQ2-A} support prioritizing requirements for AK management approaches, and especially about which ADD steps to support better. For instance, researchers should give higher priority to extending AK capturing approaches for the ``Identify design concepts'' and the ``Instantiate architecture elements'' steps, as Google has lower effectiveness in supporting these steps. This is probably because Google cannot determine the context of software engineers when identifying or instantiating architectural solutions. To support the ``Identify design concepts'' step, researchers could propose approaches that relate business requirements, design issues and alternative architectural solutions. Moreover, to support the ``Instantiate architecture elements'' step, design decisions and their rationale should be captured from existing systems and shared with practitioners.

The results in Section \ref{sec:RQ2.2} verify that blogs and tutorials are valid options for exploring and capturing AK, because they are the most relevant and highly ranked. However, during our web pages classification (see Section \ref{sec:classification}), we found that blogs and tutorials are quite diverse (e.g. private versus company blogs). This makes it very challenging to apply information retrieval or extraction techniques on blogs. Thus, we propose first exploring the different types of blogs and tutorials and the AK concepts inside them. Moreover, the results in Section \ref{sec:RQ2.2} can support developing specialized software architecture searching approaches. One idea is to re-rank results of Google differently for each ADD step; this can be achieved by developing heuristics based on the effectiveness of AK sources to support ADD steps (see Fig. \ref{fig:AKsourcesNDCG}).

\subsection {RQ4: AK concepts in highly relevant web pages}

\subsubsection{Implications for practitioners}
Section \ref{sec:RelevanceConcepts} shows that solution drawbacks and made design decisions are the most frequently appearing AK concepts within highly relevant web pages. This indicates the importance of solution drawbacks and made design decisions for practitioners. On the one hand, solution drawbacks provide AK on when an architectural solution could be discouraged. This is important for practitioners to prevent selecting architectural solutions, which must be replaced later due to their drawbacks. Replacing architectural solutions after their implementation often requires substantial effort. On the other hand, made design decisions can be potentially reused by practitioners.
Thus, practitioners should share more AK about solution drawbacks and their made design decisions on the web (e.g. in blogs or white papers); this can be done in academic or industry conferences, or by creating dedicated community websites for this type of AK.


\subsubsection{Implications for researchers}
The results in Section \ref{sec:RelevanceConcepts} support prioritizing AK management approaches to focus on important AK concepts with the highest relevance to practitioners. For example, researchers should consider automatically capturing AK related to made design decisions and solution drawbacks more than other AK concepts, because they contain the most relevant AK for making design decisions. Developing such approaches can be challenging because both made design decisions and solutions drawbacks present minorities compared to other AK concepts on the web (see Fig. \ref{fig:ConceptsPerScource}). Thus, using classification algorithms could be useful to filter web pages on the web with drawbacks or decisions.


\section{Threats to validity}
\label{sec:threats}

\subsubsection{Construct validity}
In our study in Section \ref{sec:execution}, the experience and background of participants (see Table \ref{tab:practioners}) might have influenced their assessment of web pages (i.e. specifying the relevance and the AK concepts). To mitigate this, the participants received training, and materials (as a user guide and video). 
In addition, the participants were accompanied by the researchers at the beginning of the study. Also, since participants solved some tasks on their own time, we did not have full control of their behavior (e.g. fatigue). However, we tried to mitigate this by changing the sequence of tasks for each participant. 

Another threat of validity is the possibility of mistakes in gathering data from participants during the study. To mitigate this, the participants used a plugin, which captured their input and stored it in a database for analysis. In this way, we were able to verify the data for any possible mistakes, and validate it with the participants.






\subsubsection{Reliability}
The classification of web pages into their respective AK sources presents a threat to reliability. However, we tried to ensure consistency in the classification, by establishing categories that have the highest agreement between the first three authors of the paper. Moreover, to facilitate replicating the analysis, we provide our corpus online. 




\subsubsection{External validity}
Our study used Google as our case for a search engine, without exploring other search engines (e.g. Bing or Baidu). However, Google is the most popular search engine, and thus our results could be generalized on most users. Our limited number of tasks (six tasks in Table \ref{tab:tasks}) compared to the high variety of architectural tasks in practice is another threat to the external validity. To partially mitigate this, we have designed two tasks for each ADD step to reduce the dependency on a single task. Finally, the limited number of participants (see Table \ref{tab:practioners}) in the study is another threat to the external validity of results. However, the participants have different backgrounds and experiences, which supports, to some extent the generalizability of results.



\section{Related Work}
\label{sec:relatedwork}
We are not aware of any previous studies on web searching approaches to find AK. Thus, our study is the first to investigate using web searching to perform architectural tasks. In this section, we discuss some related work in the fields of AK, as well as studies on web searching in software engineering.

\textit{Architectural knowledge.}
Researchers in the field of AK have explored the main AK concepts, such as design decisions \cite{Jansen2005}, their types \cite{Kruchten2006}, rationale of decisions \cite{Tang2007a} and solutions alternatives \cite{Zimmermann2009}. These studies established the fundamental AK concepts, which we investigate in our presented study. However, they do not propose approaches for capturing or finding AK.

Some approaches propose catalogs and repositories of AK to facilitate structuring and sharing AK. For example, Elmalki and Zdun \cite{ElMalki2019GuidingArchitectures} modeled common types of ADDs for microservice architectures. Malakuti et al. \cite{Malakuti2018ASystems} created a catalog for the types of ADDs when designing IOT system. While they 
can guide software engineers during design space exploration, they require extra manual effort to find and codify AK.

Recent efforts on AK propose approaches to automatically capture AK from different AK sources using machine learning and information retrieval techniques. For example Gorton et al. \cite{GortonICSA2017} proposed an approach to identify AK in technology documentation, and specially identify documents with certain architectural tactics (as one architectural solution). Bhat et al. \cite{Bhat2017AutomaticApproach} captured AK from issue tracking systems. They especially captured and classified the different types of design decisions (as one AK concept) in issue tracking systems. Soliman et al. \cite{Soliman2018ImprovingCommunities} improved the effectiveness of searching for AK in Stackoverflow. The approach uses machine learning and heuristics to re-rank the results of search engines. However, all three approaches do not study the effectiveness of web searching to support architectural activities.

\textit{Web searching in software engineering.}
Some approaches empirically investigated the usage of web searching by software engineers. Xia et al. \cite{Xia2017WhatWeb} empirically determined the most common and most complex software engineering tasks (e.g. searching for solutions to bugs or for third party libraries) using web search engines. Rahman et al. \cite{Rahman2018EvaluatingRetrieval} studied the effectiveness of web searching when searching for source code on the web. Their results show that searching for source code is harder than searching for other  types. Hassan et al. \cite{Hassan2020AnLogs} empirically analyzed web search queries and results related to code exceptions. Moreover, they proposed an approach to capture knowledge related to code exceptions from the web. However, these three studies do not consider finding AK to perform software architectural tasks.

Other approaches tried to improve the effectiveness of retrieving code examples from the web. For example, Wang et al. \cite{Wang2011APIExample:APIs} proposed an approach to capture code examples regarding API usage from the web. Sirres et al. \cite{Sirres2018AugmentingSearch} proposed an approach to improve the effectiveness of source code retrieval by augmenting queries with knowledge from Stackoverflow and Github. Their results improve the effectiveness of source code searching compared to Google. However, their approaches focus on finding source code rather than AK.

\section{Conclusion and Future Work}
\label{sec:conclusion}
Our main goal in this paper is to explore the retrieved architectural knowledge (AK) from the web, and the effectiveness of web search engines, when performing three Attribute-Driven Design (ADD) architectural steps. To achieve our goal, we conducted an exploratory study with software engineers, who used Google to find AK for making design decisions. Our results provide several interesting results. First, we provide an overall view on the different sources of AK and their associated AK concepts on the web. This can be useful to extend current AK capturing approaches. Second, we determined the differences in the effectiveness of Google when performing the ADD steps, which help to better understand the main capabilities of web search engines to search for AK. Finally, we identified AK concepts on the web, which provide the highest benefit for practitioners when making design decisions. This can provide a guidance for practitioners to share their AK on the web. 

The results of this study motivate us to extend current AK capturing approaches to focus on the most useful AK sources (e.g. blogs), and the most scarce and useful AK concepts (e.g. design decisions and drawbacks of solutions). Moreover, we aim to propose specialized web searching approaches to enhance the effectiveness of searching for AK.
\bibliographystyle{IEEEtran}
\bibliography{references}
\end{document}